\documentclass[notoc]{JHEP3}

\usepackage{graphicx}

\title{Non-Gaussianity from violation of slow-roll in multiple inflation}
\author{Shaun Hotchkiss \& Subir Sarkar \\
Rudolf Peierls Centre for Theoretical Physics,
University of Oxford, 1 Keble Road, Oxford, OX1 3NP, UK\\
E-mail: \email{s.hotchkiss@physics.ox.ac.uk}, \email{s.sarkar@physics.ox.ac.uk}}

\received{\today} 
\accepted{\today}

\abstract{ Multiple inflation is a model based on N=1 supergravity
  wherein there are sudden changes in the mass of the inflaton because
  it couples to `flat direction' scalar fields which undergo symmetry
  breaking phase transitions as the universe cools. The resulting
  brief violations of slow-roll evolution generate a non-gaussian
  signal which we find to be oscillatory and yielding $f_\mathrm{NL}
  \sim 5-20$. This is potentially detectable by e.g. Planck but would
  require new bispectrum estimators to do so. We also derive a
  model-independent result relating the period of oscillations of a
  phase transition during inflation to the period of oscillations in
  the primordial curvature perturbations generated by the inflaton.  }


\preprint{arXiv:0910.3373}

\begin{document}

\section{Introduction}

A major goal in modern observational cosmology is to detect any
non-gaussianity of the temperature anisotropies in the cosmic
microwave background (CMB)
\cite{Komatsu:2001rj,Bartolo:2004if,Chen:2006nt}. All single field,
slow roll, inflation models give rise to anisotropies that are
gaussian \cite{Maldacena:2002vr}. Therefore detection of a significant
non-gaussian signal would falsify such models and provide new insights
into the dynamics of inflation.

Quantifying such a signal is however not a straightforward
task. Whilst gaussianity is a well-defined property, non-gaussianity
is not and the anisotropies can, in principle, deviate from
gaussianity in many different ways \cite{Chen:2006nt}. Also, the CMB
temperature anisotropies measured by WMAP are \emph{very} close to
gaussian \cite{Komatsu:2008hk}. Therefore any measure designed to
detect non-gaussianity needs to be sensitive to a very small signal.

One measure of non-gaussianity that is particularly useful is the
three point correlation function, or `bispectrum' of the temperature
anisotropies. Gaussian random variables (and functionals of gaussian
random variables) have the property that all odd power correlation
functions are zero. This makes the bispectrum the lowest order
statistic for which \emph{any} non-zero result would indicate a
departure from gaussianity. The bispectrum contains much more
information than the power spectrum as, in general, it depends on both
scale and shape. Therefore, if a non-zero bispectrum is detected it
will also be an extremely useful statistic for constraining models of
the early universe.

\medskip However there are limitations to how much information can be
inferred from the bispectrum:

\begin{itemize}

\item Modern CMB experiments possess very large numbers of pixels of
  data. e.g. $N \simeq 10^6$ for WMAP and $N
  \simeq 10^7$ for Planck \cite{Ashdown:2006ey}. Considering that each
  higher order of correlation function will require an additional
  factor of $N$ calculations, exact determination of the bispectrum
  quickly becomes impossible. Therefore various estimators have been
  constructed \cite{Fergusson:2008ra}, but each can look only for a
  specific type of bispectrum and so might miss a signal
  different from the type being searched for.

\item Higher order correlation functions suffer more from cosmic
  variance because more information is required at each scale to
  compute them. In this work we will deal with a \emph{scale
    dependent} signal, therefore limitations due to cosmic variance
  are particularly relevant.

\end{itemize}

These problems are not necessarily insurmountable. Computing power
will only continue to get stronger, making bispectrum estimators
increasingly powerful with time. For each potential signal, estimators
can also be constructed for that specific signal, ensuring processing
time is used as efficiently as possible. The estimator can be
optimised for detecting primordial non-gaussianity while
discriminating against secondary non-gaussianities arising from
e.g. unsubtracted point sources or residuals from component separation
\cite{Munshi:2009ik}. Finally, in models that have non-trivial scale
dependence, a correlation will likely exist between the power spectrum
and bispectrum as we illustrate in this paper.

One of the first methods for quantifying non-gaussianity involved
rewriting the Newtonian potential $\Phi(x)$ as \cite{Komatsu:2001rj},
\begin{equation}
\label{fNL local}
\Phi(x) = \Phi_\mathrm{L} (x) 
+ f_\mathrm{NL} (\Phi_\mathrm{L}^2 (x) - \langle\Phi_\mathrm{L}^2 (x)\rangle),
\end{equation}
where $f_\mathrm{NL}$ is a constant, and $\Phi_\mathrm{L} (x)$ is a
gaussian variable. However, this parameterisation has a rather
specific form and does not capture other possible deviations from
gaussianity. A more general parameterisation can be obtained by
allowing $f_\mathrm{NL}$ to depend explicitly on the wavevector, ${\bf
  k}$. In terms of the adiabatic curvature perturbation, $\zeta$, the
parameterisation becomes \cite{Bartolo:2004if,Seery:2005wm},
\begin{equation}
\label{fNL general}
\zeta(x) = \zeta_\mathrm{L} (x) 
- \frac{3}{5} f_\mathrm{NL}\star(\zeta_\mathrm{L}^2 (x) 
- \langle\zeta_\mathrm{L}^2 (x)\rangle).
\end{equation}
Here, the $\star$ product is used because $f_\mathrm{NL}$ has been
allowed to depend on scale.\footnote{See Eq.(229) in
  Ref.\cite{Bartolo:2004if} for how to define this product.} The
factor of -3/5 comes from the relationship between the adiabatic
curvature and the Newtonian potential at matter domination.

For $f_\mathrm{NL}$ as defined in Eq.(\ref{fNL local}), the WMAP
5-year constraint is $-9 < f_\mathrm{NL} < 111$
\cite{Komatsu:2008hk}. This reinforces the point made already: the
primordial temperature anisotropies are \emph{very} close to
gaussian. Given that the amplitude of these anisotropies is $\Delta
T/T \sim 10^{-5}$ the non-gaussian contribution to these anisotropies
is at most $\sim10^{-8}$ (or $\sim 0.1\%$ of the overall
anisotropy). In general the expectation from inflation is that the
anisotropies should be close to gaussian, therefore this result can be
considered as a success of the inflationary paradigm.

More precisely the expectation from inflation of non-gaussianity as
parameterised by Eq.(\ref{fNL general}) depends on the specific
model. It is known that for single field, slow-roll inflation with a
canonical kinetic term and a vacuum initial state, the result is
$f_\mathrm{NL} \sim \epsilon \ll 1$ \cite{Maldacena:2002vr} where
$\epsilon$ is the usual slow-roll parameter defined later in
Eq.(\ref{defeps}). The best sensitivity expected from the Planck
satellite is to $f_\mathrm{NL}$ of $\mathcal{O}(5)$ while the
secondary contribution from post-inflationary evolution is of
$\mathcal{O}(1)$ \cite{Komatsu:2009kd}. Therefore the prediction of
the simplest inflationary toy models~\footnote{By this we mean a
  generic fine-tuned potential such as the frequently used $V (\phi) =
  m^2 \phi^2$ which has not yet been convincingly obtained from a
  physical theory, especially since inflation occurs in such models at
  $\phi > M_\mathrm{P}$.} is that there should \emph{not} be a
detection of primordial non-gaussianity in the near future.

Conversely a deviation from the simplest toy models can produce a
larger value for $f_\mathrm{NL}$. The study of these effects serves
two purposes. Firstly, from the cosmological perspective, a detection
of $f_\mathrm{NL}$ will immediately rule out single field slow-roll
inflation and focus attention on determining how inflation actually
occured. Secondly, from the perspective of inflationary model building
based on fundamental physics, as the bounds on $f_\mathrm{NL}$ grow
tighter, some interesting models can be ruled out if there is no
detection.

Much work has been done on non-gaussianity generated due to multiple
scalar fields (e.g. Ref.\cite{Sasaki:2006kq}), non-canonical kinetic
terms (e.g. Ref.\cite{Langlois:2008qf}) or non-vacuum initial states
(e.g. Ref.\cite{Martin:1999fa}). However, surprisingly little
attention has been paid to the possibility of non-gaussianity
generated by a violation of slow-roll. Ref.\cite{Byrnes:2009qy} did
consider this in a context when multiple fields are present; however,
the non-gaussianity itself is generated by the multiple fields, not
the violation of slow-roll (see also Ref.\cite{Cai:2009hw}).
Refs.\cite{Chen:2006xjb,Chen:2008wn} do consider the non-gaussianity
generated by violation of slow-roll, but for a toy inflationary
model. We follow their methods closely but consider a \emph{physical}
model of inflation. In both models the violation of slow-roll
generates sharp features in the power spectrum and also generates a
ringing in the bispectrum with a characteristic period which we
calculate below.

\subsection{Features in the Power Spectrum of Primordial Fluctuations}

It is an expectation in the simplest toy models of inflation that the
power spectrum of temperature anisotropies in the CMB should be nearly
scale-invariant \cite{Lyth:1998xn}. However this need not be the case
for physical models. If the observed spectrum was indeed perfectly
scale-invariant, such models would be ruled out and na\"ively this
might seem to be the case. Assuming the ``concordance'' $\Lambda$CDM
cosmology, the primordial spectrum is well parameterised as a power
law with spectral index $n_\mathrm{s} = 0.960 \pm 0.014$
\cite{Komatsu:2008hk}. This indicates that there cannot be any
significant scale dependence that affects the spectrum over a large
range of scales but it does \emph{not} preclude the existence of
e.g. localised oscillations in the spectrum, or other sharp
features. Moreover since the observed anistrotopies arise from the
convolution of the {\em unknown} primordial spectrum with the transfer
function of the {\em assumed} cosmological model whose parameters are
being determined, it is obvious that both unknowns cannot be
determined simultaneously without further assumptions.

In fact, there are indications that the primordial spectrum might not
be a scale-free power law, even assuming the $\Lambda$CDM cosmology
\cite{Martin:2003sg,Kogo:2004vt,Shafieloo:2003gf,Tocchini-Valentini:2004ht,Nicholson:2009pi,Ichiki:2009zz}. At
large angular scales ($>50^0$) there is essentially no power and there
are anomalous `glitches', especially in the range of multipoles $\ell
\simeq 20-40$ \cite{Spergel:2003cb,Hinshaw:2006ia}. The statistical
significance of these anomalies is not sufficient to claim a definite
detection, however it is strong enough to provide some tension with
the fit to a scale-free power-law primordial spectrum which has only a
3\% probability of being a good description of the WMAP 1-year data
\cite{Spergel:2003cb}, although this did improve to $\sim7\%$ with the
WMAP 3-year data release \cite{Hinshaw:2006ia}. Future measurements of
the $EE$ and $TE$ mode polarisations by the Planck and (proposed)
CMBPol satellites will throw light on whether these glitches are real
or not \cite{Mortonson:2009qv}.

{\it ``In the absence of an established theoretical framework in which
  to interpret these glitches (beyond the Gaussian, random phase
  paradigm), they will likely remain curiosities''}
\cite{Hinshaw:2006ia}. Indeed if there were a model that had
predicted, without ambiguity, the position and amplitude of the
glitches, this would be seen as very strong evidence for the
model. Although no such model exists, the general possibility of
generating glitches over a range of scales had in fact been proposed
prior to the WMAP observations in the context of `multiple inflation'
wherein the mass of the inflaton field undergoes sudden changes during
inflation \cite{Adams:1997de}. This generates characteristic localized
oscillations in the spectrum, as was demonstrated numerically in a toy
model of a inflationary potential with a `kink' parameterised as
\cite{Adams:2001vc}:
\begin{equation}
  V (\phi) = \frac{1}{2} m^2 \phi^2 
   \left[1 + c \tanh \left(\frac{\phi - \phi_\mathrm{s}}{d}\right)\right].
\label{kinkpot}
\end{equation}
By tuning the position, amplitude and gradient of the kink, the
locations of the glitches can be varied to match the glitches seen in
the power spectrum. One can then perform statistical likelihood tests
to determine whether the fit is better with the glitches, but with the
additional parameters, or with the simple scale-free spectrum. It was
found by the WMAP team that the fit to the 1-year data improves
significantly (by $\Delta\chi^2 = 10$) for the model parameters
$\phi_\mathrm{s} = 15.5\,M_\mathrm{P}$, $c = 9.1\times10^{-4}$ and $d
= 1.4\times10^{-2}\,M_\mathrm{P}$, where
$M_\mathrm{P}\equiv(8\pi\,G_\mathrm{N})^{-1/2} \simeq
2.44\times10^{18}$~GeV \cite{Peiris:2003ff}. This analysis was
repeated later using the WMAP 3-year data, with similar results
\cite{Covi:2006ci}.

This seems encouraging, however $m$ in the toy model above is {\em
  not} the mass of the inflaton --- in fact in all such monomial
`chaotic' inflation models with $V \propto \phi^n$, inflation occurs
at field values $\phi_\mathrm{infl} > M_\mathrm{P}$, hence the leading
term in a Taylor expansion of the potential around
$\phi_\mathrm{infl}$ is always linear in $\phi$ (since this is not a
point of symmetry), rather than quadratic as for a mass term
\cite{German:1999gi}. The effect of a change in the inflaton mass can
be sensibly modelled only in `new' inflation where inflation occurs at
field values $\phi_\mathrm{infl} << M_{\rm Pl}$ and an effective field
theory description of the inflaton potential is possible. The
`slow-roll' conditions are violated when the inflaton mass changes due
to its (gravitational) coupling to `flat direction' fields which
undergo thermal phase transitions as the universe cools during
inflation \cite{Adams:1997de}. The resulting effect on the spectrum of
the curvature perturbation was found by analytic solution of the
governing equations to correspond to a `step' followed by rapidly
damped oscillations \cite{Hunt:2004vt}.\footnote{Although a similar
  phenomenon had been noted earlier for the case where the inflaton
  potential has a jump in its slope \cite{Starobinsky:ts}, such
  a discontinuity has no physical interpretation.}

The next step should be to predict other observables, having used the
power spectrum to constrain all the parameters in the model. In this
paper we calculate the bispectrum of the multiple inflation model
\cite{Adams:1997de}, using its predicted power spectrum
\cite{Hunt:2004vt} and the set of parameters which provide the best
reduced $\chi^2$ in the $\Lambda$CDM cosmology \cite{Hunt:2007dn}. We
also examine the effect on the bispectrum of varying these parameters
over their full natural range.\footnote{We use the word `natural' in
  this context to mean ``stable towards radiative
  corrections''.}  The non-gaussianity is found to be potentially
detectable by the Planck satellite, or perhaps even a reanalysis of
the WMAP data.

\section{Multiple inflation}

The biggest difficulty in inflationary model building is obtaining a
potential that is both sufficiently flat to support inflation for
$\sim 60$ e-folds {\em and} stable towards radiative corrections. $N=1$
supergravity (SUGRA), the locally realised version of supersymmetry
(SUSY), is a natural framework for achieving this (see
Ref.\cite{Chung:2003fi} for a comprehensive review, especially of the
cosmological issues discussed below). In SUSY there are usually many
`flat directions' in field space, i.e.  scalar degrees of freedom
which, while SUSY remains unbroken, have perfectly flat
potentials. When SUSY is broken, the flat directions are lifted and
usually acquire a mass-squared related to the SUSY breaking scale. It
would seem natural to identify one of these flat directions as the
inflaton but to achieve a sufficient number of e-folds of inflation,
the mass-squared of the inflaton needs to be much smaller than the
Hubble expansion rate during inflation --- if it is not, the evolution
is too quick and inflation ends after only a few e-folds. However, due
to SUSY breaking by the vacuum energy driving inflation, the natural
expectation is $m_\phi^2 \sim H^2$. This is commonly known as the
$\eta$ problem in supergravity and superstring model building because
it relates to the slow-roll parameter $\eta$ (defined in
Eq.\ref{defeta}) being too large to support inflation.

There has been much work done to attempt to circumvent the $\eta$
problem (see Ref.\cite{Randall:1997kx} for a review and
Ref.\cite{Kachru:2003sx,Easson:2009kk} for some recent attempts in the
framework of string/brane models). In this work we will assume that
the inflaton mass has a symmetry protecting it against SUSY-breaking
corrections. This can be done for example by having the inflaton in a
`hidden sector' where it interacts with the other sectors only
gravitationally \cite{Ross:1995dq}.

In the simplest inflationary models, scalar fields other than the
inflaton itself are usually ignored. This is justified by the argument
that, because of their larger masses, they will not contribute to the
adiabatic density perturbations produced during inflation. If these
fields decay before the end of inflation then their decay products
will be diluted by the expansion of the universe and they will also
not contribute any isocurvature perturbations. Such fields usually
cannot couple to the inflaton either because any such coupling could
endanger the flatness of the inflaton potential.

Multiple inflation \cite{Adams:1997de} is based on $N=1$ SUGRA and
assumes that while `flat direction' scalar fields have their natural
SUSY-breaking scale masses of order the Hubble parameter, there is
just one scalar field (the inflaton) with a mass that is kept
anomalously small through a symmetry. The amplitude of the potential
during inflation will be the SUSY breaking scale (which need not be
the electroweak scale as SUSY breaking can be different during and after
inflation). The flat directions couple gravitationally to the inflaton
and thus affect its evolution when their own masses change as the
universe cools during inflation and they undergo symmetry-breaking
phase transitions.

\subsection{Evolution of the flat direction fields}

It is assumed \cite{Adams:1997de} that before inflation begins the
universe is in a thermal state at temperature $T$. At this point the
flat directions, $\psi$, are trapped at the origin by a thermal
potential $\propto \psi^2 T^2$. Unless protected by a symmetry, the
mass $\mu$ of the field will be of the order of the Hubble parameter,
$H$, which is itself determined by the scale $V_0 \sim \Delta^4$ of
the vacuum energy driving inflation. Henceforth we work in units where
the reduced Planck mass $M_\mathrm{P} \doteq 1$, therefore $\mu^2
\simeq H^2$ and $H^2 \simeq \Delta^2/3$.

Although the mass-squared is {\em negative}, the flat directions will be
lifted at large field values by non-renormalisable operators $\propto
\psi^n$ ($M_\mathrm{P} = 1$). For small field values, at non-zero
temperatures, the full potential for the flat directions is thus
\cite{Adams:1997de},
\begin{equation}
  V(\psi) = V_0 + \left (-\frac{\mu^2}{2} + \alpha T^2 \right) \psi^2 
   + \gamma \psi^n.
\label{flatdir}
\end{equation}
Before inflation, when the temperature is much larger than the mass of
the field, it is trapped at the origin by a thermal barrier (quantum
tunneling through which is negligible). When inflation starts, the
temperature drops rapidly and after $\sim \ln(M_\mathrm{P}/\Delta)
\approx 10$ e-folds of inflationary expansion the thermal barrier
disappears. The field can then evolve to its minimum at $\Sigma =
(\mu^2/n \gamma)^{1/n-2}$ and meanwhile there are $\sim \ln(\Sigma
M_\mathrm{P}/\Delta^2)$ e-folds more of inflation
\cite{Adams:1997de}.\footnote{The field evolves as $\psi \propto
  e^{Ht}$ so the occupation number of thermal states drops as $\sim
  e^{-e^{Ht}}$. There is no non-gaussianity generated due to the
  thermal fluctuations while the field is at the origin ({\em c.f.}
  \cite{Das:2009sg}) since observable perturbations today leave the
  horizon much later.} The field evolves (as a critically damped
oscillator) according to the equation of motion:
\begin{equation}
   \ddot{\psi} + 3H \dot{\psi} = -\frac{\partial V}{\partial \psi}
      = \left[\mu^2 - n \gamma \psi^{(n-2)} \right] \psi;
\end{equation}
when it reaches the minimum the oscillations of the field are damped
rapidly (within 1--2 cycles) by the $3 H \dot{\psi}$ term (see Fig.1
of \cite{Hunt:2004vt}). Damping due to particle creation is far less
effective since this is happening in an {\em inflating} background
\cite{Adams:1997de}. Consequently no isocurvature pertubations are
generated by such oscillations.

\subsection*{The relationship of the flat directions to the inflaton}

It was shown in Ref.\cite{Adams:1997de} that a term $\kappa \phi^\dagger
\phi \psi^2$ in the K\"{a}hler potential with $\kappa$ of ${\cal
  O}(1)$ is consistent with the underlying symmetries and corresponds
to a term $\frac{\lambda}{2} \phi^2 \psi^2$ in the scalar potential
where $\lambda = \kappa H^2$. The full potential, including the
inflaton, is thus:
\begin{equation}
V (\phi, \psi_i) = V_0 - \frac{1}{2} m_\phi^2 \phi^2 
 - \sum_i \left(\frac{1}{2}\mu_i^2\psi_i^2 
 - \frac{1}{2} \lambda_i \phi^2 \psi_i^2 
 - \gamma \psi_i^{n_i} \right).
\label{multpot}
\end{equation}
There will be a running of the inflaton mass too, however the field
value of the inflaton remains small throughout inflation so we can
neglect the higher-dimensional operators that would parameterise such
running.

Hence as each field, $\psi_i$, rolls to its minimum at $\Sigma_i$, the
effective inflaton mass-squared changes by $\lambda_i \Sigma_i^2$. The
amplitude of the potential also changes as each field falls into its
minimum, however, this change will be very small compared to the
dominant term in the potential $V_0$ \cite{Adams:1997de}.

During periods of slow-roll, the amplitude of the primordial
perturbation spectrum can be expressed in terms of the potential and
its derivatives as,
\begin{equation}
\label{powspect}
P_k = \frac{H_*^2}{8 \pi^2 \epsilon_*} = \frac{V^3}{12 \pi^2 {V'}^2}.
\end{equation}
Where the $*$ indicates that the quantities should be evaluated as the
relevant scales `cross the horizon', i.e. when $k^2=2a^2H^2$. A
transition in the mass of the inflaton, though not causing a
significant change in $V$, will cause a significant change in $V'$. If
$\lambda_i$ is positive, the mass will decrease causing a jump in the
amplitude of the spectrum, while if it is negative then the amplitude
will fall. As the flat direction oscillates in its minimum, the mass
of the coupled inflaton will briefly oscillate as well. If the
amplitude of the oscillations is large enough, one would expect a
`ringing' in the primordial power spectrum \cite{Hunt:2007dn} for
which there is tentative observational indication
\cite{Shafieloo:2003gf,Tocchini-Valentini:2004ht,Nicholson:2009pi,Ichiki:2009zz}.

The WMAP measurements indicate that, from $\ell \simeq 40$ through to
the limit of $\ell \sim 800$ set by signal-to-noise considerations,
the inferred primordial spectrum is nearly flat. However, as was noted
earlier, there is tentative evidence of a step at very large scales of
order the present Hubble radius
\cite{Martin:2003sg,Shafieloo:2003gf,Tocchini-Valentini:2004ht}. It
was in fact a prediction of the original paper on multiple inflation
\cite{Adams:1997de} that there should be of ${\cal O}(1)$ such feature
in the $\sim 10$ e-folds of inflationary history probed by the CMB.

The advantage of the multiple inflation model over the kink potential
(\ref{kinkpot}) considered in Ref.\cite{Chen:2006xjb,Adams:2001vc} (or
the toy model considered in Ref.\cite{Starobinsky:ts}) is its
grounding in a consistent, particle physics framework. While the
latter model is able to improve the fit to the glitches in the data
\cite{Peiris:2003ff,Covi:2006ci} the constraints derived on the model
parameters do not tell us anything about the relevant physical
processes at these high energy scales. By contrast, constraints on
multiple inflation are related directly to the masses and couplings of
fields potentially present during inflation. For example, the scalar
potential of the Minimal Supersymmetric Standard Model (MSSM) is flat
along many directions in field space and a catalogue exists of all
these directions \cite{Gherghetta:1995dv}. These would be lifted
during inflation due to SUSY-breaking by the large vacuum energy
present and conceivably undergo symmetry breaking phase transitions.

\section{Method}

\subsection{Single field justification}

We follow closely the method of Ref.\cite{Chen:2006xjb}, who use the
framework of Ref.\cite{Maldacena:2002vr} in calculating the
bispectrum. This method uses the comoving gauge where, in single field
inflation, the inflaton fluctuations are zero at all orders. When
there are other, more massive, fields present this is no longer true
because these extra fields will slightly perturb the instantaneous,
adiabatic, direction in field space away from the direction of the
main inflaton field. In multiple field scenarios a simpler gauge to
describe the evolution of perturbations within the horizon is the
uniform curvature gauge. In this gauge, neglecting gravitational
waves, the slicing of spacetime into equal time hypersurfaces is made
such that the spatial curvature on each of these surfaces is zero.

It was shown in Ref.\cite{Sasaki:1995aw} (and generalised beyond
linear order in Ref.\cite{Lyth:2004gb}) that there exists a
gauge-independent, non-perturbative quantity $\zeta$ which, on
sufficiently large scales, coincides with the adiabatic, scalar
density perturbation at the perturbative level. Defining the number of
e-folds of expansion as $N = \int{H \mathrm{d}t}$, it was shown in
Ref.\cite{Sasaki:1995aw,Lyth:2004gb} that if we start in the uniform
curvature gauge, this quantity can be expressed as
$\zeta=\mathrm{d}N$. This allows us (on sufficiently large scales) to
express the adiabatic curvature perturbation as \cite{Lyth:2005fi},
\begin{equation}
\label{deltaN}
\zeta = \sum_I {\frac{\partial N}{\partial \phi^I} \delta\phi^I} 
 + \sum_{I,J}{\frac{1}{2} \frac{\partial^2 N}{\partial \phi^I \partial \phi^J} 
    \delta \phi^I \delta \phi^J} + \cdots .
\end{equation}
If we set the initial, uniform curvature, hypersurface a few e-folds
of expansion after the modes cross the horizon then due to the much
smaller mass of the inflaton, we have $\dot{\phi} \ll \dot{\psi}$,
hence $\partial N/\partial \phi \ (= H/\dot{\phi}) \gg \partial
N/\partial \psi$ for all the flat direction fields. For all fields
involved (including the inflaton), the higher order terms in
Eq.(\ref{deltaN}) are also very small.\footnote{These terms become
  significant in models where non-gaussianity is seeded by entropy
  perturbations, after the perturbations have left the horizon
  \cite{Lyth:2002my}.} Under these conditions, the result $\zeta =
\frac{\partial N}{\partial \phi} \delta\phi$ holds to a very good
approximation. The flat direction fields do not affect the adiabatic
curvature perturbation and thus isocurvature perturbations do not
arise gravitationally during inflation. Therefore the inflaton
fluctuations will be zero in the comoving gauge during inflation and
we are free to use the action derived in Ref.\cite{Maldacena:2002vr}
using this gauge.

It should be stressed however that the flat directions \emph{do} play
a crucial role in the generation of the non-gaussianities. The point
of the preceding argument is that they do this only through
their coupling to the inflaton and not directly through their coupling
to gravity. What this means for the calculation is that when solving
the zeroth order equations of motion for the inflaton we must
include the effects of the flat direction fields. However, when
computing the perturbative action to quadratic and cubic order in the
perturbations, we need consider only the inflaton.

In principle it might have been less confusing to work initially in
the uniform curvature gauge, where the contribution from all the
fields is unambiguous, and to convert to the comoving gauge once the
curvature perturbation has left the horizon. This would have the
benefit of less ambiguity in the present case and be a necessity in
genuine multi-field models. In future work we intend to generalise the
method of Ref.\cite{Chen:2006xjb}, as used in this paper, to
inflationary models involving several, dynamically equally important,
fields.

\subsection{Calculating the bispectrum}

We are interested in calculating the bispectrum (three-point
correlation function) of the quantum observable $\zeta$, corresponding
to the scalar, adiabatic curvature perturbation. The model we are
considering has the standard, purely gaussian vacuum state, and the
quadratic action of the field, $\zeta$, is effectively just the free
field action (i.e. a kinetic term and a mass term). A free gaussian
field will remain gaussian, therefore in order to probe any
non-gaussianity in the fluctuations it is necessary to consider the
cubic action. We know from observations of the power spectrum that
$|\zeta| \sim 10^{-5}$ i.e. higher terms in the perturbative expansion
of the action will be less important in general.\footnote{This need
  not be true always, e.g. if a symmetry forces the cubic action to be
  anomalously small then the quartic action will produce the dominant
  contribution to any non-gaussianity \cite{Engel:2008fu}.} Therefore,
we need only consider the action up to cubic order.

We proceed by re-writing the quantum operator $\zeta (t)$ in the
interaction picture:
\begin{equation}
\zeta_\mathrm{I} (t) = \mathrm{e}^{iH_\mathrm{f} (t - t_0)}\zeta_\mathrm{I} (t_0) 
 \mathrm{e}^{-iH_\mathrm{f} (t-t_0)},
\end{equation}
where $H_\mathrm{f}$ is the free field Hamiltonian derivable from the
quadratic action and $\zeta_\mathrm{I} (t_0)$ is the initial vacuum
state value of the operator. The full quantum operator is then,
\begin{equation}
\zeta (t) = U_\mathrm{I} (t, t_0)^\dagger \zeta_\mathrm{I} (t) 
U_\mathrm{I} (t, t_0) ,
\end{equation}
where,
\begin{equation}
U_\mathrm{I} (t, t_0)=T \mathrm{e}^{-i\int^t_{t_0} H_\mathrm{I} (t') \mathrm{d}t'} ,
\end{equation}
is the time evolution operator for the interaction terms (and $T$
denotes the time ordering operator).

Putting these terms into the three-point function gives,
\begin{eqnarray}
\label{bispcomm}
\nonumber 
\langle \zeta(\mathbf{k_1}) \zeta(\mathbf{k_2}) \zeta(\mathbf{k_3}) \rangle 
&=& \langle U_\mathrm{I}^\dagger \zeta_\mathrm{I} (t)^3 U_\mathrm{I} \rangle \\
&=& -i\int^t_{t_0} \langle [\zeta_\mathrm{I} (t)^3, H_\mathrm{I} (t')] \rangle 
\mathrm{d}t' .
\end{eqnarray}
Where the last line of equality holds to first order in the expansion
of the exponentials.

It is usually emphasised at this point that the vacuum with respect to
which the expectation value is being taken is the fully interacting
vacuum, \emph{not} the free vacuum. This is important because the
$\zeta_\mathrm{I}$ appearing in these equations will be expressed in
terms of the ladder operators that annihilate the free
vacuum. Fortunately, this is not an issue in this particular case
because, similarly to what is done in Minkowski space, one can
slightly deform the integral contour into Euclidean space. This
projects out the free vacuum, as explicitly shown in
Ref.\cite{Seery:2005wm}.

Up to field redefinitions proportional to the free field equations of
motion, the Hamiltonian in Eq.(\ref{bispcomm}) can be obtained
from the action derived in Ref.\cite{Maldacena:2002vr} (Eq.(3.9) 
in that paper and Eq.(3.4) in Ref.\cite{Chen:2006xjb}). This
action can be expressed as a sum of terms whose coefficients are
functions of the slow-roll parameters, defined as:
\begin{equation}
\label{defeps} 
\epsilon \equiv \frac{\dot{\phi}^2}{2H^2} ,
\end{equation}
\begin{equation} 
\label{defeta}
  \eta \equiv \frac{\dot{\epsilon}}{\epsilon H} \simeq
  \frac{2 \ddot{\phi}}{\dot{\phi} H} +
  \frac{\dot{\phi}^2}{H^2} ,
\end{equation}
where the approximate equality holds when both $\epsilon$ and $\eta$
are small. Note that there are other common definitions for $\epsilon$
and $\eta$ in terms of derivatives of the potential: $\epsilon_V
\equiv \frac{1}{2} (V'/V)^2$ and $\eta_V = V''/V$; the ones defined
above are related through $\epsilon_V = \epsilon$ and $\eta_V =
-\frac{1}{2} \eta + 2\epsilon$. These equalities hold in the slow-roll
approximation when all four quantities are small.

In terms of these quantities, the cubic action for $\zeta$ is:,
\begin{eqnarray}
\label{cubeaction}
\nonumber 
S_3 = & \int
  \mathrm{d}^4x & a^3 \epsilon^2 \zeta \dot{\zeta}^2+ a\epsilon^2 \zeta
  (\partial \zeta)^2 - 2a^3\epsilon \dot{\zeta}(\partial
  \zeta)(\partial \chi) + \frac{a^3 \epsilon}{2} 
  \frac{\mathrm{d}\eta}{\mathrm{d}t}
  \zeta^2 \dot{\zeta} \\ &&+ \frac{a^3 \epsilon}{2}
  (\partial\zeta)(\partial \chi)\partial^2 \chi+\frac{a^3 \epsilon}{4}
  (\partial^2 \zeta)(\partial \chi)^2 + 2 f(\zeta) \left
    . \frac{\delta L}{\delta \zeta}\right |_1, 
\end{eqnarray}
where $\chi = \epsilon \partial^{-2} \dot{\zeta}$.

The term $2 f(\zeta) \left . \frac{\delta L}{\delta \zeta}\right |_1$
is proportional to the free field equations of motion, $\left
  . \frac{\delta L}{\delta \zeta}\right |_1$, and in
Ref.\cite{Maldacena:2002vr} is removed by a field redefinition. This
can have an important effect on the bispectrum, however, in the
present case it does not, due to its relative size. There is just one
term in $f(\zeta)$ (Eq.(3.10) in Ref.\cite{Maldacena:2002vr}) which
does not include a derivative on one of the $\zeta$ terms. As we can
choose to apply the correction due to the redefinition \emph{after}
the relevant scales have left the horizon, only this term will have a
non-zero effect --- its coefficient is $\eta/4$, which is much smaller
than our leading term.\footnote{Note that although $\eta$ does become
  large during the phase transition in multiple inflation, we can
  choose to apply the correction after the phase transition is over,
  at which point $\eta$ is again $\ll 1$. The redefined field will
  evolve outside the horizon while $\eta$ is significant so this can
  be done only if we also calculate the evolution of the redefined
  bispectrum out to this time, which we have indeed done.}

With the field redefinition taken care of, the Hamiltonian in
Eq.(\ref{bispcomm}) just becomes $H_\mathrm{I} = -L_3$. To use this to
evaluate Eq.(\ref{bispcomm}) we need to solve the commutator in this
equation. We are free to use the standard commutation relations for
$\zeta_\mathrm{I}$; to do so we first need to express
$\zeta_\mathrm{I} (\mathbf{k},t)$ in terms of the ladder operators of
the free field vacuum
\begin{equation} 
\zeta_\mathrm{I} (\mathbf{k}, t) = u_{\bf k} (t) a_{\bf k} +
  u^*_{\bf -k} (t) a^\dagger_{\bf -k}, 
\end{equation}
with the usual commutation relations, $[a_{\bf k}, a^\dagger_{\bf
    k'}] = (2\pi)^3 \delta^3({\bf k} - {\bf k'})$ and mode functions
$u_{\bf k} (t)$. The relationship between $\zeta_\mathrm{I} ({\bf k})$ and
$\zeta_I({\bf x})$ is the usual
\begin{equation} 
\zeta_\mathrm{I}(\mathbf{x}) = \int \frac{\mathrm{d}^3k}{(2\pi)^3}
  \zeta_\mathrm{I} (\mathbf{k}) \mathrm{e}^{i {\bf k\cdot x}}. 
\end{equation}
When these definitions are substituted back into Eq.(\ref{bispcomm}),
the result is proportional to an integral over time of the free field
mode functions $u_{\bf k} (t)$, evaluated on the interaction
Hamiltonian, $H_\mathrm{I}$ (\ref{bisp}).

\subsubsection*{The dominant term in multiple inflation}

It has been stated already that multiple inflation violates slow-roll,
nevertheless $\epsilon$ remains $\ll 1$ throughout multiple inflation
for all parameter values considered here.\footnote{If $\epsilon$ does
  become large, inflation will stop; this is potentially possible if
  there is a later inflationary epoch \cite{Adams:1997de} but we do
  not consider this possibility here.} There is only one term in the action
(\ref{cubeaction}) that does not have a factor of at least
$\epsilon^2$ in it --- this is the term with
$\mathrm{d}\eta/\mathrm{d}t$ (note that $\chi$ is of order $\epsilon$)
.  However although $\epsilon$ remains small, due to the transition in
the inflaton mass, $\eta$ and $\mathrm{d}\eta/\mathrm{d}t$ can
temporarily become large. The size of the latter term is dictated by
both the magnitude and the rate of the mass change in the phase
transition. Multiple inflation is a small field model, hence
$\epsilon$ is very small ($< 10^{-10}$) and the
$\mathrm{d}\eta/\mathrm{d}t$ term dominates over all others because it
has one less factor of $\epsilon$.

\subsubsection{Numerical method for calculating bispectrum}

Upon substituting this leading term into Eq.(\ref{bispcomm}),
using the definitions of $\zeta({\bf k})$ and $\zeta({\bf x})$ and
working through the commutation, the result is (in conformal time
$\mathrm{d}\tau \equiv \mathrm{d}t/a$):
\begin{eqnarray}
\label{bisp}
\nonumber 
\langle \zeta(\mathbf{k_1}, \tau)
\zeta(\mathbf{k_2}, \tau) \zeta(\mathbf{k_3}, \tau) \rangle 
= -2 \,\mathrm{Im}\left \{u_{{\bf k}_1} (\tau) u_{{\bf k}_2} (\tau)u_{{\bf k}_3} (\tau)
  \int^{\tau}_{\tau_0} \mathrm{d}\tau' \, \left[a^2 \epsilon 
    \frac{\mathrm{d}\eta}{\mathrm{d}\tau'} (2\pi)^3 \delta^3 \left (\sum_i {\bf k}_i \right) 
    \times \right . \right . \\ \left . \left . \left (
      u^*_{{\bf k}_1} (\tau')u^*_{{\bf k}_2} (\tau') \frac{\mathrm{d} u^*_{{\bf  k}_3}}{\mathrm{d}\tau'} 
      + u^*_{{\bf k}_1} (\tau')u^*_{{\bf k}_3} (\tau') \frac{\mathrm{d} u^*_{{\bf k}_2}}{\mathrm{d}\tau'} 
      + u^*_{{\bf k}_2} (\tau')u^*_{{\bf k}_3} (\tau') \frac{\mathrm{d} u^*_{{\bf k}_1}}{\mathrm{d}\tau'}\right) \right] \right\}.
\end{eqnarray}
This should be compared to Eq.(3.21) in Ref.\cite{Chen:2006xjb}.

To obtain the bispectrum for multiple inflation, we must solve this
integral. To do this we first need to numerically solve for the free
field mode functions $u_{\bf k}(\tau)$, the scale factor $a(\tau)$ and
the slow roll parameters $\epsilon$ and
$\mathrm{d}\eta/\mathrm{d}\tau$. Once this has been done, the results
are put into the above integral and evaluated out to a value of $\tau$
after the end of inflation. All that is needed after that is a
rescaling of the full bispectrum to obtain a scale dependent
generalisation of $f_\mathrm{NL}$.

Fortunately, the slow roll parameters in the above integral are
composite variables of other variables more fundamental to the model
(e.g. $\epsilon = \dot{\phi}^2/2H^2$). Therefore, we need only solve
for these more fundamental variables and then evaluate $\epsilon$ and
$\mathrm{d}\eta/\mathrm{d}t$ from them. The full list of variables
that need to be solved for, with their equations of motion, are
\cite{Mukhanov:1990me}:
\begin{equation}
\label{modefunc} 
v_k''+ \left(k^2- \frac{z''}{z} \right ) v_k = 0,
\end{equation}
with $z = a\sqrt{2\epsilon}$, $v_k = z u_k$,
and:
\begin{eqnarray}
\phi'' + a^2 \frac{\mathrm{d}V}{\mathrm{d}\phi} +\frac{2a'\phi'}{a} = 0,  \\
  \psi'' + a^2 \frac{\mathrm{d}V}{\mathrm{d}\psi} + \frac{2a'\psi'}{a} = 0,\\
  a'' + \frac{a}{6}\left (\phi'^2 +\psi'^2 \right) - \frac{2a^3 V}{3} = 0.
\end{eqnarray}
In all of the above, the $'$ indicates differentiation with respect to
conformal time. These four equations are just the free field equation
of motion for the observable $\zeta$, the two zeroth order
equations of motion for $\phi$ and $\psi$, and finally the Friedmann
equation for the Hubble parameter $\dot{a}/a$.

We solve these equations numerically, using the {\tt Matlab} function ode113
\cite{MatOde:1997}. For initial conditions we take the Bunch-Davies
vacuum state \cite{Mukhanov:1990me}. This amounts to:
\begin{equation} 
v_k (\tau_0) = \sqrt{\frac{1}{2k}},
\end{equation}
\begin{equation} 
v_k' (\tau_0) = -i\sqrt{\frac{k}{2}}.
\end{equation}
The initial value of $\tau = \tau_0$ in the integrand was set such that
integration of the mode functions, $v_k$, begins when the following
condition is first satisfied:
$$10^4 \frac{z''}{z} > k^2.$$
At earlier times, the mode functions are highly oscillatory and would
cancel in the integral in Eq.(\ref{bisp}). In fact there is only a
small window in which this integral is non-zero --- very early on, the
mode functions are highly oscillatory and any contribution to this
integral will be washed out, while very late on, the mode functions
are frozen and thus the $\mathrm{d}u_k/\mathrm{d}\tau$ term will be
zero. Finally, when the phase transition is not occurring, the
$\mathrm{d}\eta/\mathrm{d}\tau$ term will be negligible. It is only
those modes that are leaving the horizon during the phase transition
that will give a a non-zero.contribution to the bispectrum.

A regularisation scheme is used in Ref.\cite{Chen:2006xjb} to counter
a sharp cutoff in the integral in Eq.(\ref{bisp}) (and improved upon
in Ref.\cite{Chen:2008wn}). We find that a regularisation scheme is
useful for calculational efficiency but not necessary for accuracy. It
was argued in Ref.\cite{Chen:2006xjb} that although the integral is
technically convergent as $\tau \rightarrow -\infty$, this sharp
cutoff would add a spurious contribution to the bispectrum, due to the
highly oscillatory nature of the mode functions at early times. We
find that this is a problem only when the cutoff also co-incides with
the violation of slow-roll. If this occurs, there was indeed a
spurious contribution due to the sharp nature of the cutoff; however,
if we begin our integration earlier, when the slow-roll contribution
to the integral, $\epsilon \mathrm{d}\eta/\mathrm{d}\tau$ is
negligible, no spurious contribution is obtained.

In ordinary slow-roll, to accurately calculate the contribution from
the other, dominant terms in Eq.(\ref{cubeaction}) by numerical means,
a regularisation scheme would be necessary. For multiple inflation
(and for the kink model considered in Ref.\cite{Chen:2006xjb}) these
other terms are small enough to be ignored. This is especially true in
multiple inflation where $\epsilon$ is exceptionally small. What the
regularisation scheme does do is allow the integration to begin later
without sacrificing accuracy. Nevertheless, we did not use a
regularisation in calculating our results and altered the integration
start time when there was a risk of a spurious contribution.

\subsubsection{Comparing a bispectrum to $f_\mathrm{NL}$}

We have defined a scale-dependent generalisation of the
$f_\mathrm{NL}$ parameter in Eq.(\ref{fNL general}). Now we discuss
how to calculate $f_\mathrm{NL}$ from a given bispectrum. For a flat
bispectrum a convention has been well established, however there is an
ambiguity in the present case due to the scale dependence of both the
power spectrum and the bispectrum. For a flat spectrum, following
Refs.\cite{Maldacena:2002vr,Seery:2005wm}, we start by re-expressing
the bispectrum as,
\begin{eqnarray}
\label{bispec}
\langle \zeta(\mathbf{k_1}) \zeta(\mathbf{k_2}) \zeta(\mathbf{k_3}) \rangle 
&=& (2\pi)^3 \delta^3 \left ( \sum_i {\bf k}_i \right ) 
\frac{H^4}{16 \epsilon^2}
\frac{\mathcal{A}}{\Pi_i k_i^3}.
\end{eqnarray}
In this notation, we can rewrite $f_\mathrm{NL}$ in the
form~\footnote{This follows directly from Eq.(\ref{fNL general}) if we
  rewrite it in $k$ space and then evaluate the bispectrum.}
\begin{equation}
\label{f_NL} 
f_\mathrm{NL} = -\frac{5}{6}
  \frac{\mathcal{A}}{\sum_i k_i^3}.
\end{equation}
However, as pointed out in Ref.\cite{Chen:2006xjb}, this definition
entangles some of the ringing of the power spectrum in the definition
of $f_\mathrm{NL}$. The factors of $H^2/\epsilon$ in the bispectrum
that are divided out to get $\mathcal{A}$ are proportional to the
power spectrum which is itself oscillatory. This can be overcome 
by factoring out a number $(\tilde{P_k})^2$ that is equal to the square
of the value of the power spectrum that is measured when 
assuming scale invariance \cite{Chen:2006xjb}. This gives
\begin{equation}
\langle \zeta(\mathbf{k_1}) \zeta(\mathbf{k_2}) \zeta(\mathbf{k_3}) \rangle
 = (2 \pi)^7 \delta^3 \left( 
    \sum_i {\bf k}_i \right )\left ( \frac{\tilde{P}_k}{2}\right )^2 
 \frac{\mathcal{A}}{\Pi_i k_i^3}.
\end{equation}
In the case of slow-roll and a flat spectrum this definition is
equivalent to our previous one in Eq.(\ref{bispec}) (with $\tilde{P}_k =
H^2/8\pi^2\epsilon$). It is then appropriate to define $f_\mathrm{NL}$
in the same way as in Eq.(\ref{f_NL}) and it is this that we discuss
in the following section.  Note that although
Refs.\cite{Maldacena:2002vr,Seery:2005wm,Chen:2006xjb} define
$\mathcal{A}$ differently, the definitions for $f_\mathrm{NL}$
coincide. We have followed the convention of
Ref.\cite{Seery:2005wm} for $\mathcal{A}$.

\section{Results}

\subsection{Reproduction of the kink model}

To check the accuracy of our code and calculations we first set out to
reproduce the bispectrum of the `kink' model in Eq.(\ref{kinkpot})
\cite{Chen:2006xjb}. Note that $\phi_\mathrm{s}$ dictates where the
kink occurs, $c$ dictates the magnitude of the total shift in slope
and $d$ dictates the rate of the change in slope. For models where the
non-gaussianity is generated as the modes cross the horizon, the
bispectrum will be peaked over the equilateral shape (${\bf k}_1 =
{\bf k}_2= {\bf k}_3$).  The authors of Ref.\cite{Chen:2006xjb}
consider the statistic $\mathcal{G}(k_1,k_2,k_3)/k_1 k_2 k_3$, where
$f_\mathrm{NL} = -\frac{10k_1k_2k_3}{3 \sum_i k_i^3}
\frac{\mathcal{G}(k_1,k_2,k_3)}{k_1k_2k_3}$; in the equilateral case
this becomes $f_\mathrm{NL} = -\frac{10}{9}\frac{\mathcal{G}}{k^3}$.
In Fig.\ref{kinkequilatfull} we show the equilateral bispectrum for
the kink model over the full range over which it is non-zero, to be
compared with Fig.4 of Ref.\cite{Chen:2006xjb} which shows this over a
narrower range in $k$. Note that although $\mathcal{G}/k^3
\sim -f_\mathrm{NL}$ becomes as large as 10, it does not remain large
over an extensive range of $k$. This has important consequences for
any attempt to detect the non-gaussianity.
\FIGURE{
\includegraphics[width=0.66\textwidth]{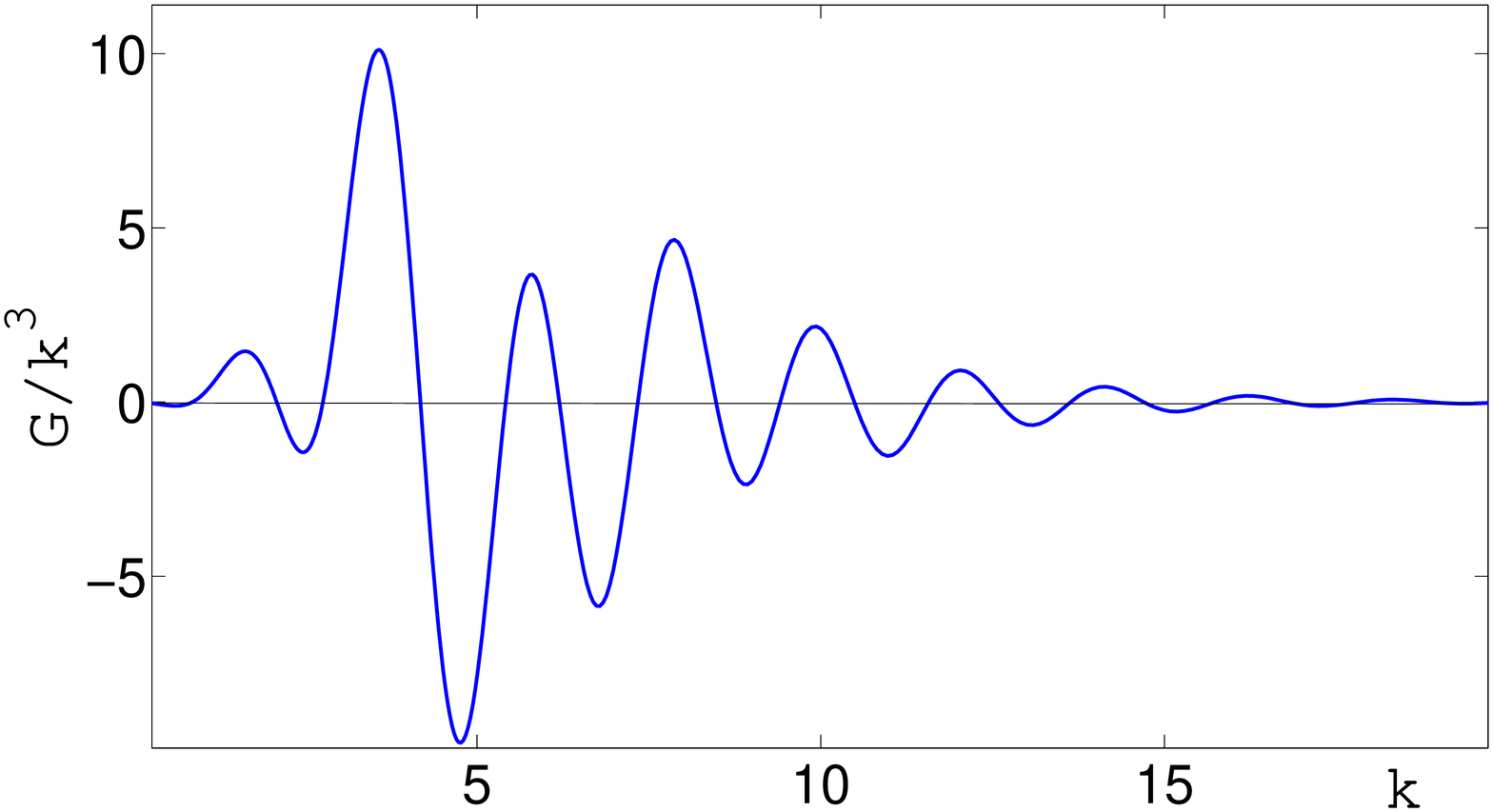}
\caption{Full equilateral bispectrum for the `kink' model
  \cite{Chen:2006xjb}, with $m=10^{-6}$, $c=0.0018$, $\phi_s = 14.81$
  and $d=0.022$ \cite{Covi:2006ci} ($M_\mathrm{P}=1$). The scale of
  $k$ is arbitrary.}
\label{kinkequilatfull}
}

\subsection{Bispectrum of multiple inflation}

The most important, and difficult, part of the bispectrum
calculation is obtaining the mode functions, $u_k$. Therefore we first
demonstrate that we can reproduce the power spectrum for multiple inflation  
accurately as shown in
Fig.\ref{multpower} (to be compared with Fig.2 of Ref.\cite{Hunt:2007dn}).
\FIGURE{
\includegraphics[width=0.66\textwidth]{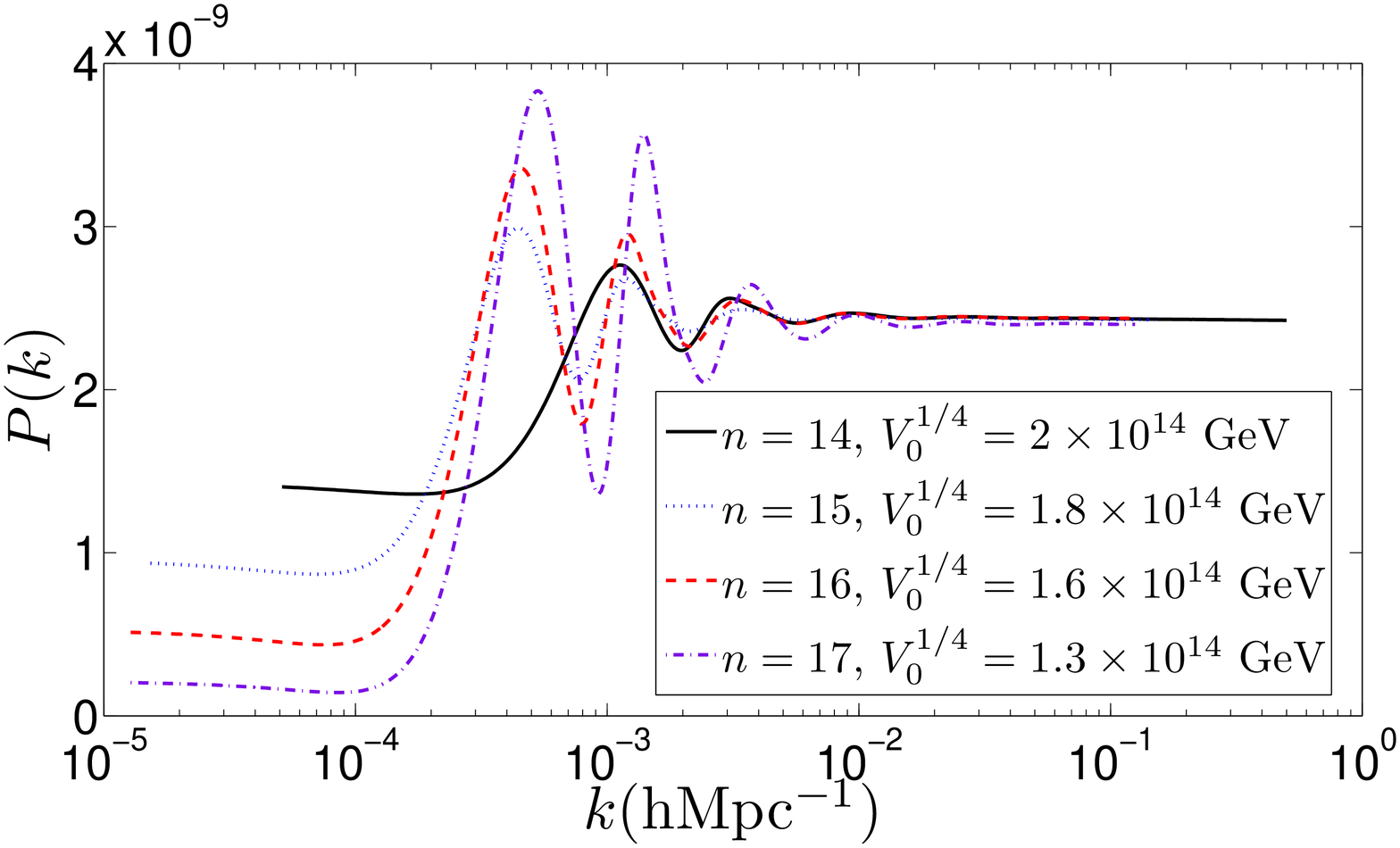}
\caption{The power spectrum for multiple inflation \cite{Hunt:2007dn},
  reproduced for various values of the order $n$ of the
  non-renormalisable operator in the potential (\ref{flatdir}). The
  inflation scale is chosen correspondingly so as to yield the same
  amplitude for the density perturbation on CMB scales.}
\label{multpower}
}
These spectra were calculated considering only one flat direction
field, with a positive coupling ($\lambda >0$) to the inflaton. The
parameters used in the multiple inflation potential (\ref{multpot})
are $\phi_0 = 0.01$, $m^2 = 0.005H^2$, $\lambda = H^2$, $\gamma=1$ and
$\mu^2 = 3H^2$ \cite{Hunt:2007dn}. Note that $V_0$ is fixed through
Eq.(\ref{powspect}) by requiring the amplitude of perturbations to be
the observed value ($\simeq 2.5\times10^{-9}$) after the phase
transition, and is different for each value of $n$, the order of the
non-renormalisable operator which lifts the flat direction potential
at large values of the field. The time at which the phase transition
occurs also varies with $n$. We emphasise that the value $n=16$ is
particularly favoured as it is a well known flat direction in the MSSM
scalar potential \cite{Gherghetta:1995dv}.

We can now confidently use our code to calculate the bispectrum of
multiple inflation as shown in Fig.\ref{bisphunt}. Note that as $k
\rightarrow 10^{-2}\,h\,\mathrm{Mpc}^{-1}$, the bispectrum begins to
exhibit noise --- this is a consequence of our not applying the
regularisation scheme described in Ref.\cite{Chen:2006xjb} and is only
present at these scales for the reasons explained earlier. Secondly,
the non-gaussianity, while not as large as in the kink model, is
non-zero over a wider range of $k$. This can be traced to the
fact that being a physical rather than a toy model, the features in
the power spectrum generated by multiple inflation \cite{Hunt:2007dn}
are not as sharp as in the kink model \cite{Adams:2001vc}. Because the
dominant term in the action is proportional to
$\mathrm{d}\eta/\mathrm{d}t$, if the transition is not sharp, then the
contribution to the bispectrum will not be large either. In multiple
inflation the $\psi$ field evolves to its minimum analogous to a
critically damped oscillator and oscillates for a period determined by
the Hubble damping, which means that the bispectrum is non-zero for
longer.
\FIGURE{
\includegraphics[width=0.66\textwidth]{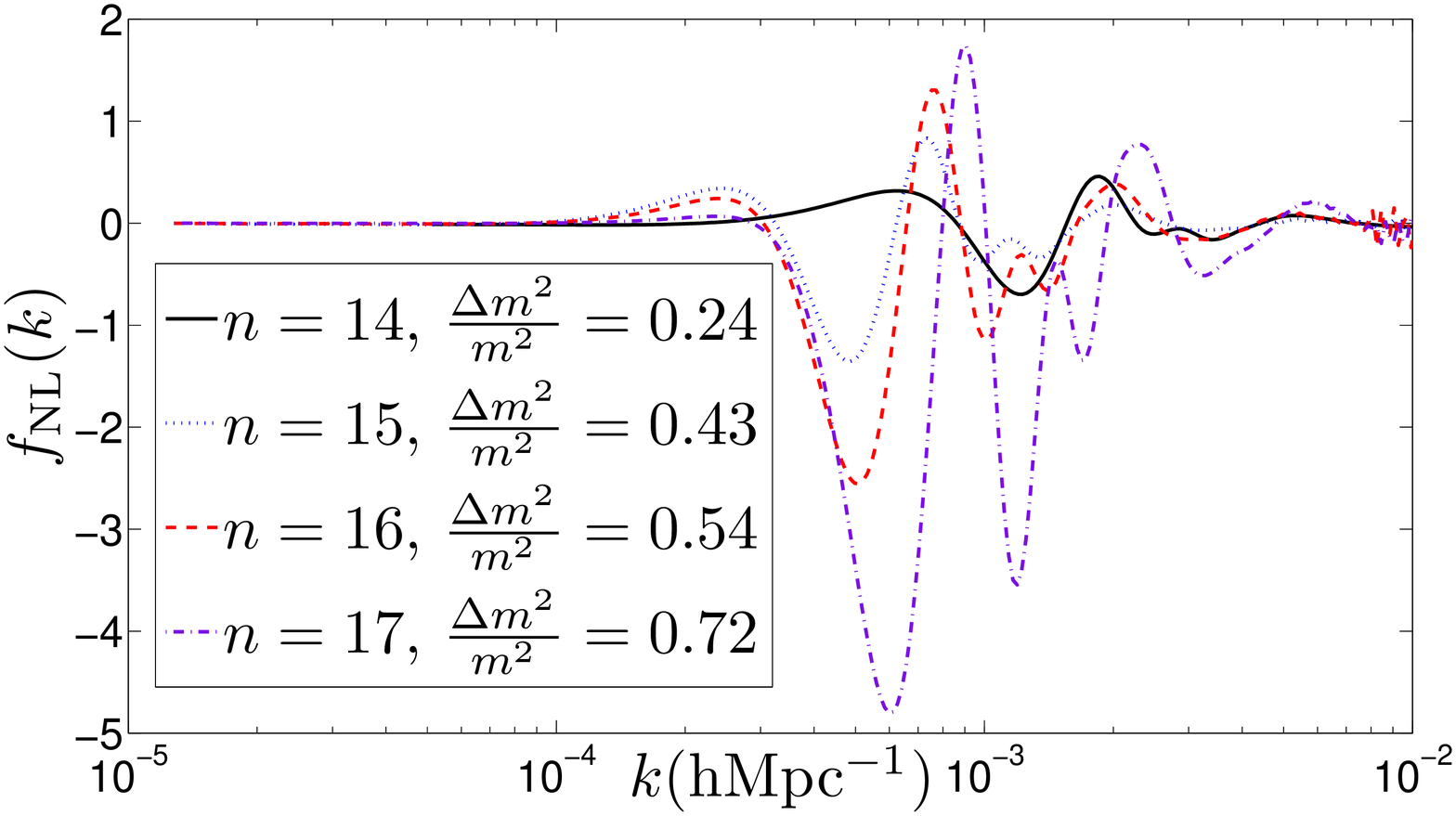}
\caption{$f_\mathrm{NL}$ for the bispectrum of multiple inflation,
  with parameters that fit the WMAP 3-yr $TT$ and $EE$ power spectra
  \cite{Hunt:2007dn}.}
\label{bisphunt}
}

Due to cosmic variance (the transition occurs at large scales) and
secondary non-gaussianities, it would be difficult to detect this
signal even with a perfect estimator. We have a similar concern about
the bispectrum of the kink model \cite{Chen:2006xjb} because although
it has a larger amplitude, it is non-zero over only a decade in $k$
making it very hard to acquire sufficient statistical significance for
a convincing detection.

\subsubsection{The effect of varying parameters}

In Ref.\cite{Hunt:2007dn}, when finding the best-fit for multiple
inflation to the WMAP $TT$ and $EE$ data, the only parameters varied
were the amplitude of the potential, the location of the phase
transition and the power $n$ of the non-renormalisable term lifting
the flat direction. All other parameters were held fixed at their most
\emph{natural} values, in order to maximise the predictive power of
the model. Unlike the toy kink model \cite{Chen:2006xjb} or other
empirical models \cite{Joy:2007na}, the parameters of multiple
inflation are not free to range over any set of values because of its
grounding in the effective field theory framework. If one simply
allowed all parameters to vary and then performed a na\"ive $\chi^2$
test, the physically constrained (and thus predictive) nature of
multiple inflation would not be factored in.

We note in passing that there is a tension between the two methods of
determining the likelihood of a model being correct. From the
fundamental physics perspective one can calculate the parameter values
that give the best fit and then estimate how natural they are. From
the cosmology model fitting perspective one can determine the goodness
of fit, given the number of parameters needed to achieve it. Neither
test is fully satisfactory because natural valued parameters could
give a bad fit and a good fit could come from an unnatural model. A
Bayesian treatment that can quantify the naturalness of a model as a
prior probability distribution would solve this problem. The resulting
Bayesian likelihood would incorporate both the naturalness of the
underlying model and the goodness of the fit to the data.

Irrespective of this issue, if multiple inflation is true, we do not
know for certain what the values of the parameters will be. Therefore
we have explored the effect of varying the parameters $\lambda$,
$\mu^2$ and $\gamma$ in turn and show the results in
Figs. \ref{varlam}, \ref{varmusq} and \ref{vargam}. In each plot, we
change $V_0$ as we vary the corresponding parameter to ensure that the
power spectrum is always normalised to its observed value of
$2.5\times 10^{-9}$. For ease of comparison we also alter the starting
point of the phase transition so that the oscillations in each figure
begin at the same point (this does not alter the shape of the
bispectrum in any way). Each plot has been approximately equated to
spatial scales probed by the CMB observations, although we are not
here comparing with actual data.
\FIGURE{
\includegraphics[width=0.66\textwidth]{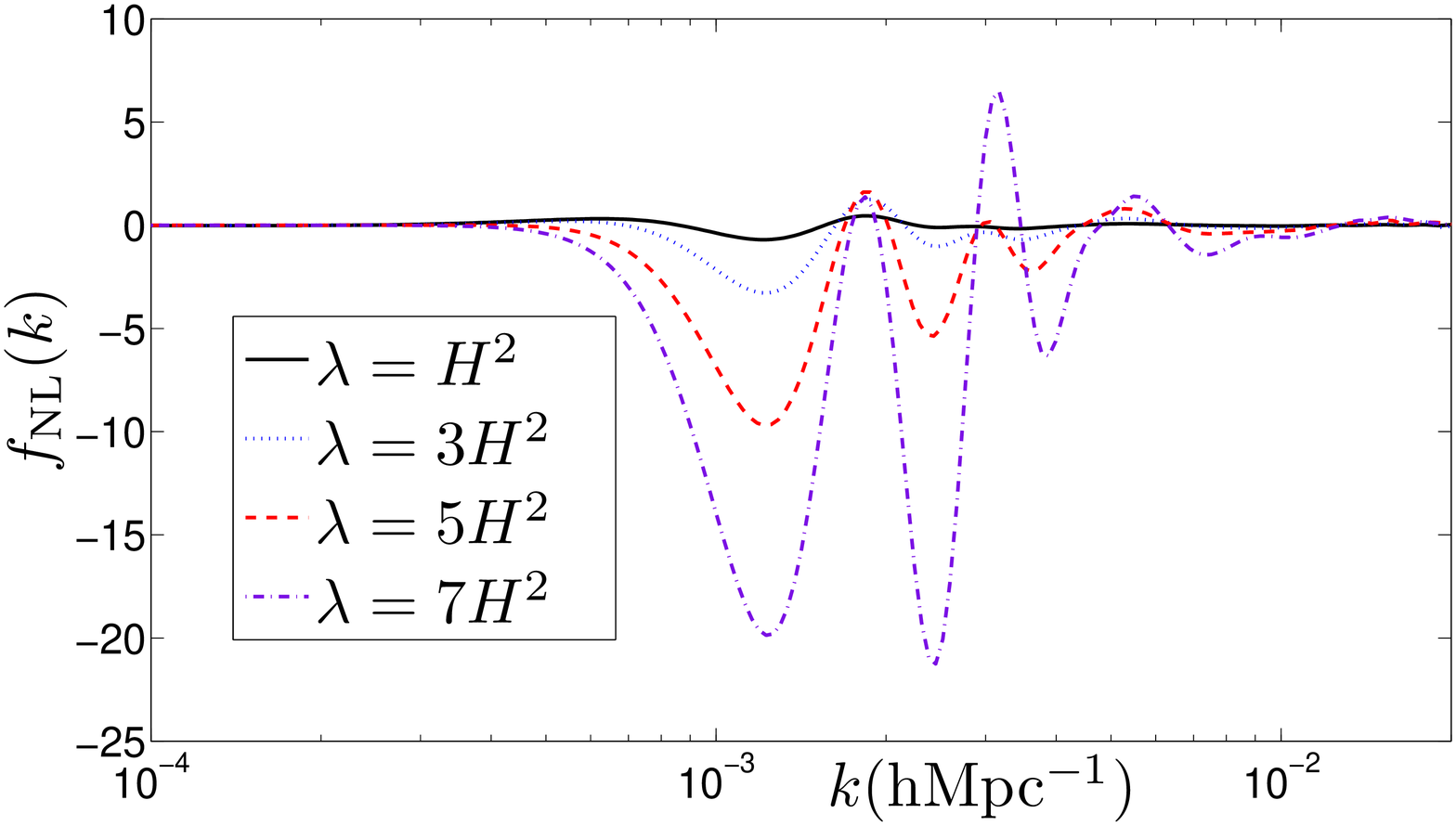}
\caption{Effect of parameter $\lambda$ on the bispectrum ($n=14$,
  $\gamma=1$ and $\mu^2=3 H^2$ for each curve).}
\label{varlam}
}

The simplest parameter to consider is $\lambda$ which does not affect
the $\psi$ field's dynamics but changes the effective mass of the
inflaton through $-m_\mathrm{eff}^2 = -m^2 + \lambda \psi^2$, so
increasing $\lambda$ increases the magnitude of the
oscillations. Therefore we expect that increasing $\lambda$ has no
effect on the bispectrum except to increase its amplitude and this is
just what we see in Fig.\ref{varlam}. It is clear that a large
bispectrum can be generated from natural values of the parameter
$\lambda$ (although values above $\sim 5 H^2$ can probably be ruled
out already from the power spectrum).

\subsubsection{The relationship between $T$ and $\ln{k}_T$}

To consider the effects of $\mu^2$ and $\gamma$ on the bispectrum it
is necessary first to derive an identity relating the period of the
oscillations over time during the phase transition to the period of
the oscillations over $k$ in the power spectrum and
bispectrum. Eq.(\ref{modefunc}) describes the time evolution of the
mode functions $u_k = v_k/z$ of the operator $\zeta_I({\bf k}, t)$. In
the slow-roll approximation, $z''/z = 2(aH)^2$. When this quantity is
small, $v_k$ oscillates as $v_k = \mathrm{e}^{ik(\tau -
  \tau_0)}/\sqrt{2k}$, and when this quantity is large we have
$v_k=\alpha z$, with $\alpha$ constant. To a good approximation, the
value of the constant $\alpha$ is determined by equating the two
solutions when $z''/z = k^2$. This approximation will continue to hold
even if $z''/z \neq 2(aH)^2$ so long as $z''/z$ continues to grow
exponentially.

The quantity $z = a\sqrt{2\epsilon}$ can be rewritten as $z =
\phi'/H$. Since $H$ is almost constant we can write $z' = \phi''/H$
and $z'' = \phi''/H$. If there are oscillations induced on $\phi'$
they will be seen more sensitively by higher derivatives of this
quantity, hence $z'' = \phi'''/H$ will have a relative amplitude of
oscillation much greater than $z' = \phi'/H$. This means that $z''/z$
will oscillate with the same frequency as $\phi'$, which in turn will
oscillate with the same frequency as $\psi$ in its minimum.

In the slow-roll limit, $a$ grows as $\mathrm{e}^{Ht}$ and this holds
even during the phase transition since the vacuum energy $V_0$ remains
sensibly constant.\footnote{This will of course not hold precisely,
  however the relative deviation will be much smaller than the
  deviation of $z''/z$ from its slow-roll value, due to the $\phi'$
  term in $z$.} Therefore, if the oscillation repeats itself every $T$
units of time, then it will repeat itself every $(\ln{a})_T = HT$
units of $\ln{a}$. It was established earlier that this means $z''/z$
will also repeat itself every $(\ln{a})_T$ units of $\ln{a}$. We can
thus parameterise the deviation of $z''/z$ from $2(aH)^2$ by the
equation:
\begin{equation} 
\frac{z''}{z} = 2(aH)^2 f(\ln{a}),
\end{equation}
where $f(x)$ is a function with period $HT$. When we do the
matching of the solutions inside and outside the horizon and equate
$\frac{z''}{z}=k^2$, we get,
\begin{equation} 
\left ( \frac{z''}{z}\right )^{\frac{1}{2}} = k = \sqrt{2}aH f^{\frac{1}{2}}.
\end{equation}
If the period of $f(x)$ is $HT$ then the period of $f^{1/2}$ will be
$2HT$ because it will repeat itself half as often. If we ask what the
difference in $k$ will be in the matching solution when $f^{1/2}$
undergoes one period, from $t=t_2$ to $t=t_1$, with $t_2-t_1=2T$, we
get,
\begin{equation} 
  k_2 - k_1 = \sqrt{2} H \left[
    \mathrm{e}^{2Ht_2} f (2Ht_2)^{1/2} - \mathrm{e}^{2Ht_1} f (2Ht_1)^{1/2}
    \right].
\end{equation}
 Unfortunately this result is not independent of the values of $t_2$
and $t_1$ and will vary as the oscillations occur. However if we
consider what will happen to $\ln{k}$ during this same time period
we get, with $f_n = f (2Ht_n)$:
\begin{equation} 
\ln{k}_2 - \ln{k}_1 = 2H(t_2 - t_1) + \frac{1}{2}(\ln{f_2} - \ln{f_1}).
\end{equation}
Now the only time dependence comes in the difference between the
function $\ln{f}$ at the two repeating points. These points were
chosen because they were at the same point of the oscillation in
$f^{1/2}$, therefore to a good approximation $\ln{f}$ will be
equal at each point. This approximation will only hold exactly if the
oscillations in $z''/z$ are exactly repeating.

It is thus expected that the mode functions $u_k$ will oscillate with
a period over $\ln{k}$ equal to $(\ln{k})_T = 2HT$, with a small drift
due to the change of the function $f(\ln{a})$ during each
oscillation. This result is general to any oscillatory phase
transition during inflation and is not specific to multiple inflation.
\FIGURE{
\includegraphics[width=0.66\textwidth]{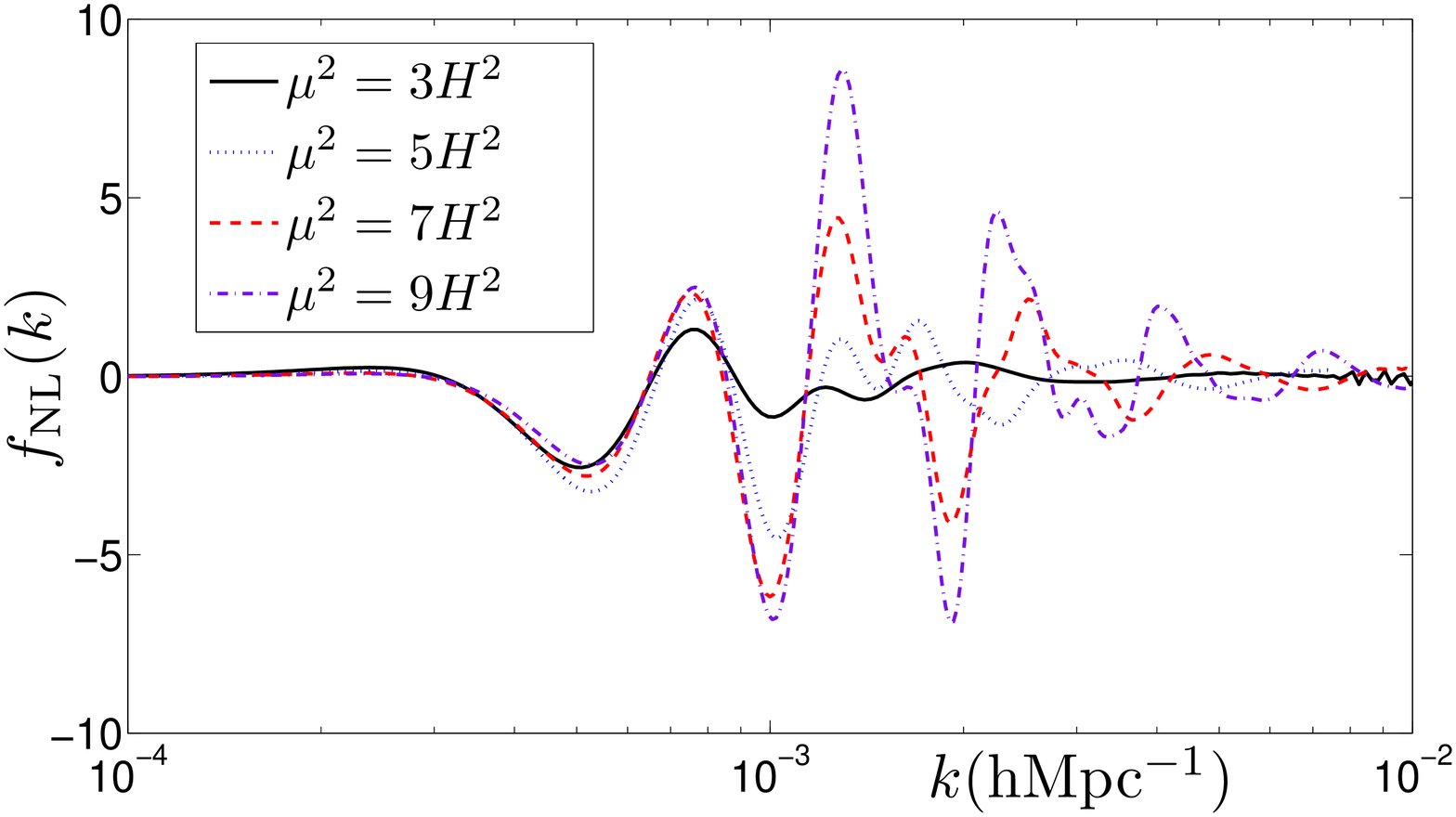}
\caption{Effect of parameter $\mu^2$ on the bispectrum ($n=16$,
  $\gamma=1$ and $\lambda=H^2$ for each curve).}
\label{varmusq}
}
It is a simple matter now to calculate the expected period, $T$, of
the oscillations in multiple inflation. The minimum of the potential
for $\psi$ occurs at the point, $\Sigma=
(\mu^2/n\gamma)^{1/(n-2)}$. If we expand the potential around this
point, with respect to the variable $\psi_\star = \psi - \Sigma$, then
the potential becomes,
\begin{equation}
V(\psi_{\star})=\frac{\mu^2 (n-1)}{2} \psi_{\star}^2 +
  \mathrm{higher\ order\ terms.}
\end{equation}
It is useful to note that $\gamma$ drops out of the quadratic term and
only appears in the higher order terms. Seen from this perspective,
the equation of motion for the field $\psi$, when in its
minimum, is to first approximation a damped harmonic oscillator. The
period, $T$, of a damped harmonic oscillator with damping term $3H$
and potential $\mu^2 (n-1) \psi_\star^2/2$ is simply:
$$H T = \frac{2 \pi}{\sqrt{\left(\frac{\mu}{H} \right)^2 (n-1) 
- \frac{9}{4}}}.$$
From the results earlier we expect then that the period of
oscillations in both the power spectrum and bispectrum should depend
only on the ratio $(\mu/H)^2$ and the power, $n$, of the
non-renormalisable term lifting the potential of the flat direction,
$\psi$. A brief check of the period of $P(\ln{k})$ confirms the
expectation that $(\ln{k})_T = 2HT$ for the mode
functions.\footnote{$P(\ln{k})$ should have period
  $(\ln{k})_T/2$ because it involves the square of the mode
  functions $u_k$. To a good approximation, the bispectrum will have a
  period $(\ln{k})_T/3$; however the bispectrum is 
  more complicated because the $u_k$ are the mode functions of the
  $\zeta_I$, which are the \emph{free} field operators. The bispectrum
  is a correlation function between the full operators
  $\zeta (t) = U_I(t,t_0)^{\dagger}\zeta_I(t)U_I(t,t_0)$ and there will
  be additional oscillations generated by the $U_I$ terms.} This can
be read off Fig.\ref{multpower} for the
parameters listed in the figure.

Therefore we expect that the result of increasing $\mu^2 H^2$ will be
to decrease the period of oscillations in the bispectrum. Since this
means more rapid oscillations, we also expect a magnification in
$\eta'$ during the phase transition, hence a magnification in the
amplitude of the bispectrum. This is just what we see in
Fig.\ref{varmusq} (there is also a small effect on the shape of the
bispectrum).
\FIGURE{
\includegraphics[width=0.66\textwidth]{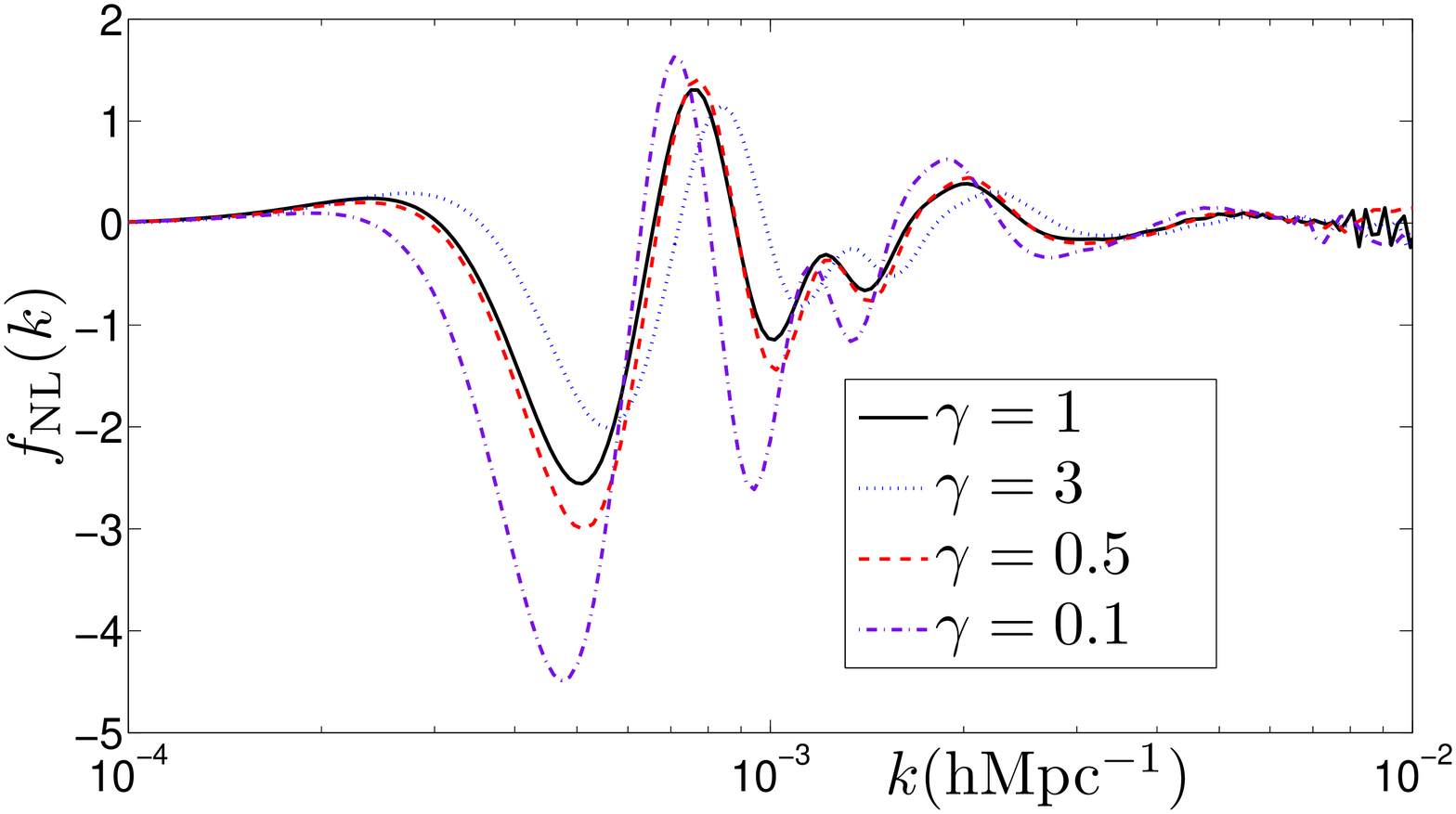}
\caption{Effect of parameter $\gamma$ on the bispectrum ($n=16$,
  $\mu^2 = 3H^2$ and $\lambda=H^2$ for each curve).}
\label{vargam}
}

Based on this calculation we expect that changing $\gamma$ will have
only a secondary effect on the bispectrum. Indeed in Fig.\ref{vargam}
the amplitude changes from $\sim 5$ to $\sim 2$ when $\gamma$ is
changed by over an order of magnitude. By comparison, when $\lambda$
is similarly altered, the change in the bispectrum amplitude was from
$\sim 1$ to $\sim 20$. The increase in the amplitude is because
changing $\gamma$ changes the minimum of the $\psi$ field, which
alters the magnitude of the change in the mass of the inflaton.

\subsection{The bump model of \cite{Hunt:2007dn}}

Finally, we point out that in Ref.\cite{Hunt:2007dn} a second example
of multiple inflation was considered with two flat directions which
have opposite sign couplings to the inflaton and produce a `bump' in
the primordial power spectrum. Using this model it was found that the
WMAP 3-year data can be fitted by an Einstein de-Sitter cosmology
without a cosmological constant. We have calculated the bispectrum for
this model which has $\mu_1^2 = \mu_2^2 = 3 H^2$,
$\gamma_1=\gamma_2=1$, $\lambda_1 = \lambda_2 = H^2$, $n_1=12$,
$n_2=13$, and find the amplitude of the oscillations induced by the
$\eta'$ term to be only $\sim 0.1$. Such a small value will be swamped
by secondary non gaussianities and could never be observed.

\section{Conclusions}

There is tentative evidence that the primordial power spectrum of
scalar perturbations is imprinted with sharp features on large scales
\cite{Spergel:2003cb,Hinshaw:2006ia}. These could have resulted from
one or more phase transitions early on in the inflationary epoch as
happens in the multiple inflation model \cite{Adams:1997de}. This
model is well motivated by fundamental physics ($N=1$ supergravity)
and such spectral features were predicted \emph{before} WMAP provided
observational indications for them.

It was shown \cite{Chen:2006xjb} that any such departure from
slow-roll during inflation should also generate non-gaussianity. We
have calculated the non-gaussianity in mutliple inflation for the
parameter values that best fit the features in the power spectrum
\cite{Hunt:2007dn}. This is on the edge of being observable but a
clear detection would require a new type of bispectrum estimator due
to its scale dependence and non factorisability
\cite{Fergusson:2008ra}. Significantly larger non-gaussianities can be
generated during multiple inflation if the parameters are allowed to
range over values which are technically natural (i.e. stable towards
radiative corrections).

The form of this non-gaussianity is tightly correlated with the power
spectrum --- the oscillations in the bispectrum should begin at the
same multipole as the oscillations in the power spectrum and have two
thirds of the period. There have been a number of attempts, to
deconvolve the primordial power spectrum directly from the WMAP data
\cite{Kogo:2004vt,Shafieloo:2003gf,Tocchini-Valentini:2004ht} (see
also Ref.\cite{Bridges:2005br}). It is generally found that there is a
supression of power on the scale of the present Hubble radius,
followed by a `ringing' at medium scales. By measuring the period of
the oscillations in the power spectrum one can predict the period (and
phase) of oscillations in the bispectrum, in a \emph{model
  independent} manner.

Forthcoming measurements of CMB polarisation by Planck ought to shed
light on whether these features in the power spectrum are systematic
errors or genuine evidence of non-trivial dynamics during the
inflationary era. The associated non-gaussian signal should also be
detectable according to our model calculation and provide insight into
the dynamics.

\subsection{Acknowledgements}

SH acknowledges a Clarendon Fellowship and support from Balliol
College, Oxford and the EU Marie Curie Network ``UniverseNet''
(HPRN-CT-2006-035863). We thank Paul Hunt, David Lyth, Graham Ross,
Misao Sasaki and David Wands for helpful comments and discussions.

\end{document}